# ALAE: Accelerating Local Alignment with Affine Gap Exactly in Biosequence Databases

Xiaochun Yang
College of Information Science
and Engineering
Northeastern University
Liaoning 110819 China
yangxc@mail.neu.edu.cn

Honglei Liu
College of Information Science
and Engineering
Northeastern University
Liaoning 110819 China
liuhonglei@gmail.com

Bin Wang
College of Information Science
and Engineering
Northeastern University
Liaoning 110819 China
binwang@mail.neu.edu.cn

## ABSTRACT

We study the problem of local alignment, which is finding pairs of similar subsequences with gaps. The problem exists in biosequence databases. BLAST is a typical software for finding local alignment based on heuristic, but could miss results. Using the Smith-Waterman algorithm, we can find all local alignments in $\mathcal{O}(mn)$ time, where $m$ and $n$ are lengths of a query and a text, respectively. A recent exact approach BWT-SW improves the complexity of the Smith-Waterman algorithm under constraints, but still much slower than BLAST. This paper takes on the challenge of designing an accurate and efficient algorithm for evaluating local-alignment searches, especially for long queries. In this paper, we propose an efficient software called ALAE to speed up BWT-SW using a compressed suffix array. ALAE utilizes a family of filtering techniques to prune meaningless calculations and an algorithm for reusing score calculations. We also give a mathematical analysis and show that the upper bound of the total number of calculated entries using ALAE could vary from $4.50mn^{0.520}$ to $9.05mn^{0.896}$ for random DNA sequences and vary from $8.28mn^{0.364}$ to $7.49mn^{0.723}$ for random protein sequences. We demonstrate the significant performance improvement of ALAE on BWT-SW using a thorough experimental study on real biosequences. ALAE guarantees correctness and accelerates BLAST for most of parameters.

## 1. INTRODUCTION

Similar to web applications, another area that has recently witnessed a rapid surge in the amount of data being produced is the biosequence search. In this area, scientists often want to compare a biosequence against a collection of known sequences. Generally, two biologically related sequences appearing dissimilar in their entirety may contain subsequences that are highly similar.

*Local alignment* is a common technique for finding a pair of highly similar substrings from two given sequences, respectively. In querying biological sequences, search tools often distinguish between short queries and long queries [9]. For example, short queries (read) are used to find the same structural or functional subunits (motifs) from very different protein families or genomes; large genomes or chromosomes, however, need to be compared in comparative genomics, such as aligning mouse genomes against human genomes [7, 12]. Generally, each biosequence can have the scale ranging from a few hundred million characters to a few billion characters and the length of a long query could be from a few thousand characters to ten million characters [7, 8]. Efficiently aligning long queries against biosequences poses a competitive challenge to the development of alignment tools.

A variety of computational algorithms have been developed for finding local alignments, among which BLAST (Basic Local Alignment Search Tool) [1, 2] and the Smith-Waterman algorithm [13] are typical ones.

BLAST [1] is a popular tool for identifying the local alignments between sequences. It decomposes an input query into a set of grams and identifies matches against the database using grams of the query. A local alignment is created by examining the left and right subsequences from these matches. Although this heuristic approach suggests a time-optimized model, it does not guarantee to find all alignment results that meet the specified score criterion.

The Smith-Waterman algorithm [13] is a well-known dynamic programming algorithm that could accurately identify the best local alignments between a query sequence and sequences in the database. It compares fragments of arbitrary lengths between two sequences and supports a flexible scoring scheme by allowing different scores for different types of operations, including substitution, insertion, and deletion. The score of an alignment is a summation of the score of each operation involved in the alignment, which makes the algorithm sensitive and ensures an optimal alignment of the sequences. However, this also has the effect that the method is very slow and CPU intensive. A recent approach called BWT-SW [8] is an exact method that improves the complexity of the Smith-Waterman algorithm under limited scoring scheme constraints, but still much slower than the very efficient approximate method BLAST.

This paper takes on the challenge of designing an efficient algorithm for evaluating local-alignment searches exactly. We improve the general dynamic programming algorithm by exploiting a family of filtering techniques and reusable calculations. The challenges and our contributions are as follows.

(1) How to avoid calculating most of entries in dynamic programming matrixes without impairing the accuracy of the alignment results? Calculating entries in matrixes is time consuming, especially when the text and query are long. We analyze the property of entries in the matrixes and propose a family of filtering techniques to avoid meaningless calculations in Section 3. Finding that there are many duplicate calculations in each matrix, we propose an algorithm for reusing those duplicates in Section 4.

(2) How to satisfy the space requirement of large biosequences for both the text and the query? We consider an in-memory algorithm and use the recent results on compressed suffix array to make our approach (called ALAE) doable in memory. The idea is similar to BWT-SW, but we adapt it to fit for our filtering techniques and reusing approaches (see Section 5).

(3) What is the upper bound of the number of calculated entries? We give a mathematical analysis and prove that ALAE could provide a time efficiency guarantee across representative ranges of user specified schemes in Section 6. The upper bound of the total number of calculated entries using ALAE could vary from $4.50mn^{0.520}$ to $9.05mn^{0.896}$ for random DNA sequences and vary from $8.28mn^{0.364}$ to $7.49mn^{0.723}$ for random proteins sequences.

In addition, in Section 7 we show experimental results on real biosequence databases including DNAs and proteins to demonstrate the space and time efficiency of our ALAE approach. We show that ALAE makes a significant improvement of performance on BWT-SW for all scoring schemes and thresholds. ALAE also accelerates BLAST for most of scoring schemes and guarantees correctness.

## 2. PRELIMINARY

Let $\Sigma$ be the alphabet of characters in biosequences. For a sequence $S$ of the characters in $\Sigma$, we use $|S|$ to denote its length, $S[i]$ to denote its $i$-th character (starting from 1), and $S[i, j]$ to denote the subsequence from its $i$-th character to its $j$-th character. The typical length of a genomic sequence is from millions to a few billion. In this section, we give the definition of local alignment.

### 2.1 Local Alignment with Affine Gap Penalty

Before formally defining local alignment, we present the widely used scoring scheme for biosequences. In this scoring scheme, each identical mapping has a positive score $s_a$, whereas a substitution of one character has a negative score $s_b$, and a gap (insertion of $r$ characters or deletion of $r$ characters) has an affine gap penalty represented as a negative score $s_g + r \times s_s$, where $s_g$ is a gap opening penalty represented as a negative score and $s_s$ is a gap extension penalty represented by another negative score for each insertion or deletion. We use $\langle s_a, s_b, s_g, s_s \rangle$ to represent a scoring scheme and use the default scoring scheme $\langle 1, -3, -5, -2 \rangle$ in both BLAST and BWT-SW to show examples in this paper, which means $s_a = 1$, $s_b = -3$, $s_g = -5$, and $s_s = -2$.

The similarity between two sequences $S_1$ and $S_2$ is defined as the value of the alignment of $S_1$ and $S_2$ that maximizes total alignment score, denoted $sim(S_1, S_2)$. For example, let $S_1$ = AAACG and $S_2$ = AACCG, then the optimal alignment of $S_1$ and $S_2$ is to replace the third character A of $S_1$ with the third character C of $S_2$, i.e. $sim(S_1, S_2) = 1 \times 4 + (-3) = 1$.

**Local alignment problem**. Let $T$ be a text sequence of $n$ characters and $P$ be a query sequence of $m$ characters. For any $1 \leq \pi_t \leq n$ and $1 \leq \pi_p \leq m$, compute the largest similarity between $T[x, \pi_t]$ and $P[y, \pi_p]$ ($1 \leq x \leq \pi_t, 1 \leq y \leq \pi_p$), i.e. the maximum alignment score of any substring of $T$ ending at position $\pi_t$ and any substring of $P$ ending at position $\pi_p$. For biological applications, we are only interested in those substring pairs if their alignment scores attain a threshold $H$[1].

One naive approach to find all of their local alignments is to examine all substrings of $T$ and align them one by one with $P$. Obviously, we want to avoid aligning $P$ with the same substring at different positions of the text $T$. A natural solution is to build a suffix trie $\mathcal{T}$ of the text $T$ as reported in [8]. Then, distinct substrings of $T$ are represented by different paths from the root to different nodes in the suffix trie. Let $p_u$ be a path from the root to a node $u$ in the suffix trie $\mathcal{T}$. We align each substring represented by $p_u$ against the query pattern $P$.

Given a data sequence $T$, a query pattern $P$, and a threshold $H$, Algorithm 1 shows the BASIC algorithm for answering local alignment using the suffix trie $\mathcal{T}$ of $T$. According to the problem definition, we use $A(i, j)$ to represent the alignment of $T[x, i]$ and $P[y, j]$ with largest alignment score ($1 \leq x \leq i, 1 \leq y \leq j$). The algorithm first initializes the largest score $A(i, j).score = 0$ for each alignment $A(i, j)$, and the starting position $A(i, j).pos = 0$ (line 1). Let $p$ be a suffix path from the root to a leaf node (line 2). For each substring $X$ represented by $p$, the BASIC algorithm searches prefix $X[1, i]$ and finds all alignment pairs $(X[1, i], P[y, j])$ whose similarities are greater than $H$ (lines 3 – 5). Each prefix $X[1, i]$ corresponds with substrings at different positions $t_1, \ldots, t_k$, which means that the alignment scores of $A(t_1 + i - 1, j)$, ..., $A(t_k + i - 1, j)$ have the same score as $sim(X[1, i], P[y, j])$. For all alignments of $X[1, i]$ and $P[y, j]$ with the same end position $t+i$ in $T$ and $j$ in $P$ ($t_1 \leq t \leq t_k$), we choose the largest alignment score among them. Let $t$ be the starting position of the alignment with largest score. The algorithm sets $A(t + i, j).pos = t$ (lines 6 – 10). It finally returns all alignments with positive scores that are greater than or equal to the threshold $H$ (line 11).

---

**Algorithm 1**: BASIC – Calculating local alignments.

**Input**: A suffix trie $\mathcal{T}$ of text $T$ with $n$ characters, a query pattern $P$ with $m$ characters, and a score threshold $H$;
**Output**: End position pairs of local alignments;
1 Initialize each alignment $A(i, j).score = 0$ and $A(i, j).pos = 0$ ($1 \leq i \leq n, 1 \leq j \leq m$);
2 **foreach** *suffix path $p$ from the root to a leaf node in $\mathcal{T}$* **do**
3     let $X$ be the same substring representing by $p$;
4     **foreach** *prefix of $X$ with $i$ characters starting at positions $t_1, \ldots, t_k$* **do**
5         align $X[1, i]$ against $P$ and find all alignment pairs such that $sim(X[1, i], P[y, j]) \geq H$ ($1 \leq y \leq j \leq m$);
6         **foreach** *above alignment pair $(X[1, i], P[y, j])$* **do**
7             **foreach** *starting position $t_h$ of $X[1, i]$ ($t_1 \leq t_h \leq t_k$)* **do**
8                 **if** $sim(X[1, i], P[y, j]) > A(t_h + i, j).score$ **then**
9                     $A(t_h + i, j).score = sim(X[1, i], P[y, j])$;
10                     $A(t_h + i, j).pos = t_h$;

11 return alignments $A(i, j)$ if $A(i, j).score \geq H$ ($1 \leq i \leq n, 1 \leq j \leq m$);

---

### 2.2 Dynamic Programming for Answering Local Alignment Exactly

Given a query pattern $P$, for each substring $X$ represented by a suffix path $p$ from the root to a leaf node, we need to align each prefix $X[1, i]$ ($1 \leq i \leq |X|$) against $P$. Let $M_X(i, j)$ be the best alignment score of $X[1, i]$ and any substring of $P$ ending at position $j$. We allow that any substring $P[y, j]$ ($1 \leq y \leq j$) is a potential match. We use an auxiliary score $G_a(i, j)$ to store the best alignment score under the restriction that $X[i]$ is aligned with a gap, and use another auxiliary score $G_b(i, j)$ to store the best alignment score under the restriction that $P[j]$ is aligned with a gap.

Initial condition:
$M_X(0, j) = 0$         for $0 \leq j \leq m$.
$M_X(i, 0) = s_g + i \times s_s$     for $1 \leq i \leq d$.
$G_a(0, j) = -\infty$         for $0 \leq j \leq m$.
$G_b(i, 0) = -\infty$         for $1 \leq i \leq d$.

---

[1]$H$ could be determined indirectly using user specified expectation value $E$-value. We discuss it in Section 7.

Recurrences (for $i > 1, j > 1$)

$$M_X(i,j) = \max \left\{ \begin{array}{l} M_X(i-1,j-1) + \delta(X[i], P[j]), \\ G_a(i,j), \\ G_b(i,j) \end{array} \right\}, \text{ where}$$

$$\delta(X[i], P[j]) = \left\{ \begin{array}{ll} s_a & \text{if } X[i] \text{ equals to } P[j], \\ s_b & \text{otherwise.} \end{array} \right.$$

$$G_a(i,j) = \max\{G_a(i-1,j) + s_s, M_X(i-1,j) + (s_g + s_s)\}.$$
$$G_b(i,j) = \max\{G_b(i,j-1) + s_s, M_X(i,j-1) + (s_g + s_s)\}.$$

|   |   |    | 1<br>G |    | 2<br>C |    | 3<br>T |    | 4<br>A |    | 5<br>G |
|---|---|----|-----|----|-----|----|-----|----|-----|----|-----|
|   |   | $-\infty$<br>**0** | | $-\infty$<br>**0** | | $-\infty$<br>**0** | | $-\infty$<br>**0** | | $-\infty$<br>**0** | | $-\infty$<br>**0** |
| 1 | G | $-\infty$ | **−7** | −14 | −7<br>**1** | −6 | −7<br>**−3** | −8 | −7<br>**−3** | −10 | −7<br>**−3** | −10 | −7<br>**1** |
| 2 | C | $-\infty$ | **−9** | −16 | −6<br>**−6** | −13 | −9<br>**2** | −5 | −9<br>**−5** | −7 | −9<br>**−6** | −7 | −6<br>**−6** |
| 3 | T | $-\infty$ | **−11** | −18 | −8<br>**−8** | −15 | −5<br>**−5** | −12 | −11<br>**3** | −4 | −11<br>**−4** | −6 | −8<br>**−6** |
| 4 | A | $-\infty$ | **−13** | −20 | −10<br>**−10** | −17 | −7<br>**−7** | −14 | −4<br>**−4** | −11 | −11<br>**4** | −3 | −10<br>**−3** |

**Figure 1: An example of calculating local alignment score (bold values represent $M_X(i,j)$).**

Fig. 1 shows an example of the dynamic programming of aligning a substring $X$=GCTA against a query $P$=GCTAG. For example, $M_X(4,3) = \max\{M_X(3,2) + \delta(X[4], P[3]), G_a(4,3), G_b(4,3)\}$. Since $X[4]$ does not equal to $P[3]$, $\delta(X[4], P[3]) = s_b = -3$. $G_a(4,3) = \max\{G_a(3,3)+s_s, M_X(3,3)+(s_g+s_s)\} = \max\{-11+(-2), 3+(-5-2)\} = -4$ and $G_b(4,3) = \max\{G_b(4,2)+s_s, M_X(4,2)+(s_g+s_s)\} = \max\{-17+(-2), -7+(-5-2)\} = -14$. Therefore, $M_X(3,4) = \max\{-5+(-3), -4, -14\} = -4$.

The variables used in this paper are shown in Table 1.

**Table 1: List of variables and their notations.**

| Variables | Notations |
|---|---|
| $T$ | A sequence (a.k.a. a text). |
| $P$ | A query pattern. |
| $n$ | Length of $T$. |
| $m$ | Length of $P$. |
| $X$ | A substring represented by a suffix path from the root to a leaf node in the suffix trie of $T$. |
| $M_X$ | A matrix for a substring $X$ of $T$ and a query $P$. |
| $M_X(i,j)$ | Best alignment score of $X[1,i]$ and $P[y,j]$, where $1 \leq y \leq j$. |
| $A(i,j)$ | The alignment with the largest alignment score of $T[x,i]$ and $P[y,j]$, where $1 \leq x \leq i, 1 \leq y \leq j$. We use $A(i,j).score$ to express the largest alignment score, and $A(i,j).pos$ to express the starting position $x$ in $T$. |

In the remaining of this paper, we will focus on local alignment on whole texts only. Notice that, the techniques can be immediately applied to collections of sequences: given all the sequences $T_1, \ldots, T_n$ in the database, we concatenate them into a single sequence $T$. A local alignment query is then performed directly on the sequence $T$.

## 2.3 Suffix Trie and Compressed Suffix Array

*Suffix Trie*. Let $T$ be a text with $n$ characters. The suffix trie of the text $T$ is a trie whose edges are labeled with strings, such that each path from the root of the trie to a leaf represents exactly one suffix of $T$. Each leaf node stores the starting location of the corresponding suffix of $T$.

*Compressed Suffix Array*. Compressed suffix array [5] is a combination of the Burrows-Wheeler compression algorithm [3] and the suffix array [10]. In [3], Burrows and Wheeler propose a new compression algorithm based on a reversible transformation, called BWT, which transforms a text $T$ into a new string that is "easy to compress." BWT appends a special symbol $ smaller than any other symbol of $\Sigma$ at the end of $T$. Let the position of $ be $n + 1$. For example, given a text $T$ = GCTAGC, we append $ at the end of $T$ and get $T'$ = GCTAGC$. Then the BWT transformation of $T'$ is CTGGA$C.

The suffix array $SA[0,n]$ of $T'$ is an array of indexes such that $SA[i]$ stores the starting position of the $i$-th lexicographically smallest suffix. For example, $SA$ of GCTAGC$ is $\{7, 4, 6, 2, 5, 1, 3\}$. The space occupancy of the compressed suffix array is optimal in an information-content sense.

The compressed suffix array can support effective searches for arbitrary patterns [5]. Given a substring $X$, we can use the backward search algorithm [6] to identify the $SA$ range of $X$ in $\mathcal{O}(|X|)$ steps. In particular, it processes the last character $c$ of $X$ in the first step. It looks at $c$ as a string $S$. Let $[i,j]$ be the $SA$ range of $S$. Then it processes the string $xS$ by iteratively inserting one character $x$ before $S$ in $X$. The backward search algorithm shows that each step could be done in constant time. For any string $X$, if there exists an $SA$ range of $X$, say $[i,j]$ in $SA$, the starting positions of $X$ in $T$ can be found in $SA[i], SA[i+1], \ldots$, and $SA[j]$. For instance, the $SA$ range of a substring GC is $[4, 5]$, then the starting positions of GC in $T$ are 5 and 1.

In Section 5 we show how to simulate traversals of a suffix trie $\mathcal{T}$ using the compressed suffix array.

## 2.4 Related Work

There are a large amount of techniques on supporting local alignments, such as [4, 11, 13, 14].

The Smith-Waterman algorithm supports slow but formally correct local-alignment searches and guarantees users the optimal local alignments between query and database sequences. It requires $\mathcal{O}(nm)$ time complexity, which is a considerable disadvantage.

OASIS [11] employs a dynamic programming A*-search which is driven by traversing a suffix tree index constructed on the database sequences. It can accurately find local alignments and outperforms both BLAST and the Smith-Waterman algorithm only when the query sequences are very short (less than 60 characters).

BWT-SW is a recently proposed exact method for finding all local alignments. It uses a BWT index to emulate the suffix trie of $T$ and modifies the dynamic programming (i.e. the BASIC algorithm) to allow pruning but without missing any results. BWT-SW traverses the suffix trie in preorder and provides an early-termination technique by ignoring all negative alignment scores. Each path from the root to an intermediate node $u$ represents multiple substrings of $T$. It shows that for any a path from the root to an intermediate node $u$, if the matrix indicates that there is not any substring of the query pattern having a positive score when aligned with the path, then BWT-SW can safely prune the subtree rooted at $u$ away. Given the fixed scoring scheme $\langle 1, -3, -5, -2 \rangle$, the expected running time is $\mathcal{O}(mn^{0.628})$ for random strings and the total number of calculated entries is upper bounded by $69mn^{0.628}$. In addition, BWT-SW requires that $|s_b| \geq 3|s_a|$, which highly limits its usability.

Although BWT-SW improves the time complexity of the Smith-Waterman algorithm under the constraint $|s_b| \geq 3|s_a|$, it is still much slower than the approximate method BLAST. By comparing the Smith-Waterman and BWT-SW algorithms with BLAST we find that BLAST gets notable differences in accuracy and speed with the former two algorithms. However, BLAST is an approximate approach that could not guarantee to find all local alignments even though BLAST is indeed accurate enough in most cases [8]. In this paper, we propose a new approach ALAE to find all alignments with a comparative speed.

## 3. AVOIDING MEANINGLESS CALCULATIONS

Ideally, we hope to only calculate the alignment scores that can generate the optimal alignments. For an entry whose alignment score is impossible to be an optimal alignment score, we call it *meaningless*, otherwise, we call it *meaningful*.

In this section, we propose a family of filtering techniques to prune meaningless entries. In Section 3.1, we show that some entries in a single matrix are meaningless and we propose *local filtering* techniques to prune those meaningless ones. In Section 3.2, we show that an entry in a matrix is meaningless if its alignment score has been calculated in some other matrixes, and we propose *global filtering* techniques across matrixes.

### 3.1 Local Filtering

We propose three local filtering techniques to prune meaningless entries in a single matrix for a substring $X$ and a query $P$.

#### 3.1.1 Length Filtering

Given a query $P$ and a score threshold $H$, the BASIC algorithm aligns each substring represented by a suffix path against $P$. In this section, we show that we only need to align substrings of $T$ with certain lengths against $P$.

THEOREM 1. *Length filtering. Given a query $P$ with $m$ characters, a substring $X$ of $T$, and a score threshold $H$. The $(i,j)$-entry of $M_X$ is meaningless if $i$ does not satisfy the following condition:*

$$\lceil \frac{H}{s_a} \rceil \leq i \leq \max\{m, m + \lfloor \frac{H - (s_a \times m + s_g)}{s_s} \rfloor\}. \quad (1)$$

*We use $L_{max}$ to express the length upper bound $\max\{m, m + \lfloor \frac{H-(s_a \times m + s_g)}{s_s} \rfloor\}$.*

PROOF. Remind that $M_X(i,j)$ is the score of aligning $X[1,i]$ against a substring $P'$ of $P$ ending at position $j$ in $P$. Let the length of the substring $P'$ be $h$.

When $i \leq h$, there are at most $i$ matches and the maximum possible score of $M_X(i,j)$ is $s_a \times i$. Since we are only interested in $M_X(i,j) \geq H$, we get $s_a \times i \geq M_X(i,j) \geq H$, i.e., $\lceil \frac{H}{s_a} \rceil \leq i \leq h$.

When $i > h$, there are at most $h$ matches and at least $i - h$ gaps. Therefore, the maximum possible score of $M_X(i,j)$ is $s_a \times h + s_g + s_s \times (i - h)$. As we have mentioned above, we are only interested in $M_X(i,j) \geq H$, we get $s_a \times h + s_g + s_s \times (i - h) \geq M_X(i,j) \geq H$. Since $s_s < 0$, $s_a - s_s > 0$, and $h \leq m$, we get $i \leq \frac{s_a + (s_a - s_s) \times m - H}{|s_s|}$, i.e., $h < i \leq m + \lfloor \frac{H-(s_a \times m + s_g)}{s_s} \rfloor$.

Accordingly, $i$ should be either in the interval $[\lceil \frac{H}{s_a} \rceil, h]$ or the interval $(h, m + \lfloor \frac{H-(s_a \times m + s_g)}{s_s} \rfloor]$. Thus, Equation 1 holds. $\square$

For example, given a text $T$=CTAGCTAG, a query $P$=GCTAC, and let the threshold $H = 3$. We only need to consider substrings of $T$ with length in between 3 and 4.

#### 3.1.2 Score Filtering

We could early terminate the calculation of a score $M_X(i,j)$ if we know for any score $M_X(i'',j'')$ based on $M_X(i,j)$ ($i'' \geq i, j'' \geq j$), $M_X(i'',j'')$ is impossible to attain the threshold $H$.

THEOREM 2. *Score filtering. For any substring $X[1,i]$ starting at position $\pi_t$ in $T$ ($1 \leq \pi_t \leq n$), the $(i,j)$-entry of $M_X$ is meaningless if:*

$$M_X(i,j) \leq \max \left\{ \begin{array}{l} 0, \\ H - (m-j) \times s_a - 1, \\ H - (\min\{L_{max}, n - \pi_t\} - i) \times s_a - 1 \end{array} \right\}.$$

PROOF. We are only interested in those alignments scores $\geq H$. Let $M_X(i,j)$ be the score of $X[1,i]$ and $P[y,j]$ ($1 \leq y \leq j$).

(i) BWT-SW shows that $M_X(i,j)$ must be greater than 0. Let $M_X(i',j')$ be the score of $X[1,i']$ and $P[y,j']$ ($i' \leq i, j' \leq j$). Assume $M_X(i',j') \leq 0$, then the score of the alignment between $X[i'+1,i]$ and $P[j'+1,j]$ must be greater than or equal to $M_X(i,j)$. According to the definition of local alignment problem in Section 2, the alignment between $X[1,i]$ and $P[y,j]$ could not be the best alignment, therefore, the $(i,j)$-entry in this matrix is meaningless.

(ii) Now we consider an alignment score $M_X(i'',j'')$ in the matrix $M_X$. Let $C$ be the alignment score of $X[i+1,i'']$ and $P[j+1,j'']$, then $M_X(i'',j'') = M_X(i,j) + C \geq H$. If $M_X(i,j) \leq H - C - 1$, then the alignment of $X[1,i'']$ and $P[y,j'']$ could not be the answer, thus it is meaningless to calculate the $(i,j)$-entry.

As we know, the only way to increase an alignment score is by a match. The largest possible value of $C$ is $(j''-j) \times s_a$ or $(\min\{L_{max}, i''\} - i) \times s_a$ since there are at most $(j''-j)$ or $(\min\{L_{max}, i''\} - i)$ matches between $X[i+1,i'']$ and $P[j+1,j'']$. As we know, the maximal $j''$ is $m$ and the maximal $i''$ is $n-\pi_t$, then the upper bound of $C$ is $(m-j) \times s_a$ or $(\min\{L_{max}, n-\pi_t\} - i) \times s_a$. Therefore, when $M_X(i,j) \leq H - C - 1 \leq \max\{H-(m-j) \times s_a - 1, H-(\min\{L_{max}, n-\pi_t\} - i) \times s_a - 1\}$, the $(i,j)$-entry of $M_X$ is meaningless. $\square$

For example, given a text $T$=CTAGCTAG. Let $X$=GCTA be a substring of $T$, and let the threshold $H = 3$. Consider the matrix $M_X$ in Fig. 1. All the entries with negative scores are meaningless. The $(1,5)$-entry is meaningless, since the lower bound of the score for the 5-th column must be 3, but the calculated $M_X(1,5) = 1$. The lower bound of the scores for the 4-th row is 3. Therefore, among the 30 entries in Fig. 1, only four entries $(1,1)$, $(2,2)$, $(3,3)$, and $(4,4)$ are meaningful according to score filtering.

#### 3.1.3 Prefix Filtering

Consider a suffix path $p$ in the suffix trie $\mathcal{T}$ of the text $T$. Let $X$ be the substring represented by the path $p$. We start from the first character of $X$ and align each prefix $X[1,i]$ ($1 \leq i \leq L_{max}$) in the query $P$. According to score filtering, we are only interested in positive alignment scores. According to the scoring scheme, only an identical mapping has a positive score $s_a$. Therefore, for an alignment of $X[1,i]$ and $P[y,j]$, there must exist an integer $q$ such that $X[1,q]$ exactly matches $P[y, y+q-1]$ ($1 \leq y \leq m-q+1$) to make their alignment score large enough to counteract the effect of a mismatch or a gap. Equation 2 defines the length value $q$ according to the scoring scheme.

$$q = \lfloor \frac{\min\{|s_b|, |s_g + s_s|\}}{s_a} \rfloor + 1. \quad (2)$$

We call the substring $X[1,q]$ *q-prefix* for the suffix path $p$. Based on the observation above, we present prefix filtering in Theorem 3.

THEOREM 3. *q-Prefix filtering. Let $M_X(i,j)$ be the alignment score of $X[1,i]$ and $P[y,j]$ ($1 \leq y \leq j$). The $(i,j)$-entry of $M_X$ is meaningless if $X[1,q]$ does not match $P[y, y+q-1]$ exactly.*

Furthermore, for a substring $X$ of $T$, if we could not find an exact match between $X[1,q]$ and a substring of $P$, entries of the whole matrix for $X$ and $P$ are meaningless. For instance, consider a substring $X$=ACACAT and a query $P$=GCGTGTGA under the scoring scheme $\langle 1, -3, -5, -2 \rangle$. All entries in the matrix for $X$ and $P$ are meaningless since we could not find an exact match of $X[1,q]$ in $P$, where $q = 4$.

In order to find the exact match of $X[1, q]$ in $P$ efficiently, we build inverted lists of $q$-grams of $P$ on the fly. We decompose $P$ into a set of $q$-grams by sliding a window of length $q$ over the characters of $P$. For each $q$-gram in $P$, we generate an inverted list of its start positions in $P$. The time complexity of building inverted lists is $\mathcal{O}(m)$.

According to Theorem 3, for each starting position $\pi_p$ of $X[1, q]$ in $P$, there must exist a *fork* area that entries outside of this fork are meaningless. Fig. 2 shows the sketch of a fork. The rectangle in the figure represents a matrix $M_X$ for $X$ and $P$. Each fork in the matrix consists of three regions: an exact match region (denoted EMR), a no gap region (denoted NGR), and a gap region. We call an entry $(l, l + \pi_p - 1)$ a first gap open entry (FGOE for short) if the entry satisfies the following two conditions:

(i) $M_X(l, \pi_p + l - 1) > |s_g + s_s|$, and
(ii) for each $i < l$ and $j = \pi_p + i - 1$, $M_X(i, j) \leq |s_g + s_s|$.

An FGOE belongs to an NGR and it is a point switch from the NGR to a gap region. From the FGOE $(l, \pi_p + l - 1)$, we need to calculate another two extension entries $(l, \pi_p + l)$ and $(l + 1, \pi_p + l - 1)$. The shape of a gap region can be determined by a set of extension entries. Each extension entry is represented by a concave corner point such that its score is greater than $|s_s|$.

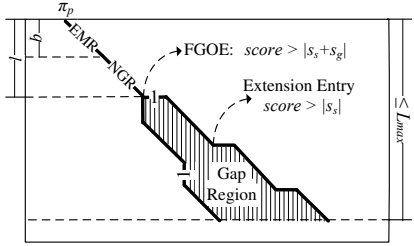

**Figure 2: Entries outside of fork areas are meaningless.**

Given a fork starting from the entry $(1, \pi_p)$, an entry $(i, j)$ belongs to its EMR if $1 \leq i \leq q$ and $j = \pi_p + i - 1$. An entry $(i, j)$ belongs to its NGR if $q < i \leq l$ and $j = \pi_p + i - 1$.

According to Theorem 3, for each entry $(i, j)$ in an EMR, its score $M_X(i, j) = i$; for each entry $(i, j)$ in an NGR, it is no need to consider the auxiliary scores in $G_a$ or $G_b$, so we can simplify the recurrence function of $M_X(i, j)$ in Section 2.2 as follows:

$$M_X(i, j) = M_X(i - 1, j - 1) + \delta(X[i], P[j]). \quad (3)$$

## 3.2 Global Filtering

Basically, given a suffix trie $\mathcal{T}$ of text $T$ with $k$ paths, we need to calculate $k$ matrixes to align a substring represented by each suffix path against the query pattern $P$. For each matrix, we need to calculate entries inside its forks. A natural question is whether we can safely avoid calculating a certain fork in a matrix based on the results of calculated matrixes. The answer to this question is yes if we could find a "good" order of calculations for suffix paths in $\mathcal{T}$.

In this section, we first analyze the effect of a calculated matrix and use bitwise operations to dynamically update and check meaningless calculations. We then show there exists dominate relationships between $q$-prefixes in $\mathcal{T}$ and observe that a fork area could be safely pruned using $q$-prefix domination. We show how to find the good order of calculations based on $q$-prefix domination and give a space-efficient approach to do global filtering for a large text.

### 3.2.1 Meaningless Fork Areas

Let $X'$ be a substring starting at position $t$ in $T$ and $X$ be the suffix of $X'$ starting at position $t+i$ in $T$. We first consider the simple case where both $X'$ and $X$ only appear once in $T$ (see Fig. 3). According to the BASIC algorithm, we need to construct two matrixes $M_{X'}$ and $M_X$. The following two cases show that calculating a fork area starting from $(1, j)$-entry in $M_X$ is meaningless.

Case (1) : $X[1, q]$ does not match $P[j, j + q - 1]$. According to our analysis in Section 3.1.3, the $(1, j)$-entry of $M_X$ is meaningless; or

Case (2) : $X[1, q]$ matches $P[j, j+q-1]$, and the alignment score $A(t+i, j).score \geq s_a$. Since there is an identity mapping between $X[1]$ and $P[j]$, $M_X(1, j)$ must be equal to $s_a$. When the matrix $M_{X'}$ makes $A(t+i, j).score \geq s_a$, we could ignore calculating $M_X(1, j)$.

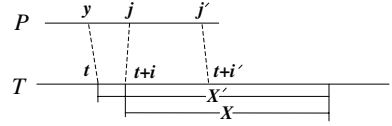

**Figure 3: For the same alignment $A(t + i', j')$, using $M_{X'}$ can produce higher score than using $M_X$ if $M_{X'}(i, j) \geq s_a$, where $M_{X'}(i, j)$ is associated with the alignment $A(t + i', j')$.**

We extend the above analysis to the general scenario that $X$ might have multiple occurrences in $T$.

THEOREM 4. *Let $X$ be represented by the suffix path $p_u$. Let $t_1, \ldots, t_k$ be the starting positions of $X$. The $(1, j)$-entry of $M_X$ is meaningless if the following two cases hold:*

*Case (1) : $X[1, q]$ does not match $P[j, j + q - 1]$; or*

*Case (2) : $X[1, q]$ matches $P[j, j + q - 1]$, and each alignment score $A(t_h, j).score \geq s_a$ $(1 \leq h \leq k)$.*

*Update and check meaningless calculations on-the-fly.* According to Theorem 4, we can avoid calculating the fork area starting from the $(1, j)$-entry in the matrix $M_X$. In order to do it, a simple way is to construct an $n \times m$ boolean matrix $G$ as follows. A $(\pi_t, \pi_p)$-entry of $G$ is 1 if $A(\pi_t, \pi_p).score \geq s_a$, otherwise, it is 0.

Consider a matrix $M_X$ and its $(1, j)$-entry. If there exists a substring $X'$ and $M_{X'}(i, j) \geq s_a$, we construct a column vector $z$. Let entries $z[t_h + i - 1]$ be 1 ($i \geq 1, t_1 \leq t_h \leq t_k$) and the remaining entries be 0. The $(1, j)$-entry of $M_X$ is meaningless if the bitwise AND operation between the $j$-th column of $G$ and $z$ equals to $z$. Then the fork area starting from this $(1, j)$-entry does not need to be calculated. If the above bitwise AND operation does not equal to $z$, we update the $j$-th column of $G$ by doing bitwise OR operation between the $j$-th column of $G$ and $z$. We repeat the above process until all suffix paths have been processed.

For example, given a text $T$=GCTAGCTA. Let $X'$=GCTA be a substring of $T$. Suppose we have calculated the matrix shown in Fig. 1 to align GCTA against the query $P$=GCTAG. Based on this matrix $M_{X'}$, we generate the following boolean matrix $G$.

$$\begin{pmatrix} 1 & 0 & 0 & 0 & 1 \\ 0 & 1 & 0 & 0 & 0 \\ 0 & 0 & 1 & 0 & 0 \\ 0 & 0 & 0 & 1 & 0 \\ 1 & 0 & 0 & 0 & 0 \\ 0 & 1 & 0 & 0 & 0 \\ 0 & 0 & 1 & 0 & 0 \\ 0 & 0 & 0 & 1 & 0 \end{pmatrix}_{8 \times 5}$$

For another substring $X$=CTAG, before constructing its matrix $M_X$, we find an exact match between $X[1, q]$=CTAG and $P[2, 5]$. We then check if the $(1, 2)$-entry of $M_X$ is meaningless. We construct a column vector $z = (0, 1, 0, 0, 0, 1, 0, 0)$ and make a bitwise

AND operation between the second column of $G$ and $z$. The result of this bitwise AND operation equals to $z$, since $X$ appears once in the text $T$ and it associates with one alignments $A(1, 1)$. Therefore, $(1, 2)$-entry of $M_X$ is meaningless.

### 3.2.2 Prune Meaningless Forks using $q$-Prefix Domination

The online approach in Section 3.2.1 requires $n \times m$ space to store the matrix $G$, which is space consuming especially when both the lengths of the text and the query are large. The question is whether there exists a "good" order of calculations to avoid using this matrix $G$. In this section, we show that the answer is yes.

Based on the analysis in Section 3.1.3, we know that each fork must be started at an exact match between the $q$-prefix of $X$ and a substring of $P$. That is, given another substring $X'$, for any fork in $M_{X'}$, there must exist at least $q$ entries with scores greater than or equal to $s_a$. We could use this property to define the order of the calculations. We formally define this property below.

DEFINITION 1. *Let $p$ and $p_1, \ldots, p_k$ be suffix paths from the root to leaf nodes in the suffix trie of $T$. Let $X^q$ and $X_1^q, \ldots, X_k^q$ be $q$-prefixes represented by $p$ and $p_1, \ldots, p_k$, respectively. If for each appearance of $X^q$ at position $t$, we can always find an appearance of a $q$-prefix in $\{X_1^q, \ldots, X_k^q\}$ at position $t - 1$, we say each $X_i^q$ $q$-dominates $X^q$, denoted $X_i^q \succ X^q$ $(1 \leq i \leq k)$.*

LEMMA 1. *Given a text $T$ and a query pattern $P$. Let $P[j, j + q - 1]$ and $P[j - 1, j + q - 2]$ be two substrings with length $q$. It is meaningless to calculate the alignment score $A(x, j)$ if one of the following conditions holds:*
- *We could not find a substring $X$ whose $q$-prefix exactly matches $P[j, j + q - 1]$;*
- *We could find two substrings $X$ and $X'$ such that $X[1, q] = P[j, j + q - 1]$, $X'[1, q] = P[j - 1, j + q - 2]$, and $X' \succ X$.*

*Constructing dominations offline.* According to Lemma 1, we need to find all dominate relationships among $q$-length substrings of the text $T$ to filter meaningless calculations. We preprocess the text $T$ and construct dominations offline in $\mathcal{O}(n)$ time as follows. We start from the first character of $T$ and scan the whole text. For any two substrings $X_1^q$ and $X_2^q$ at position $i$ and $i + 1$ with $q$-length, respectively, we construct dominate relationship $X_1^q \succ X_2^q$. We require that the $q$-length substring at position 1 could not be dominated by any other $q$-length substrings.

*Check meaningless calculations on-the-fly.* Given a query $P$, we build up $q$-gram inverted lists on-the-fly as discussed in Section 3.1.3. For each position $j$ in each gram list, we search $P[j, j + q - 1]$ and $P[j - 1, j + q - 2]$ in the text $T$ (We show a space-efficient approach to simulate traversals of the suffix trie $\mathcal{T}$ using compressed suffix array in Section 5). If we can find exact matches $X_2^q = P[j, j + q - 1]$ and $X_1^q = P[j - 1, j + q - 2]$, and $X_1^q \succ X_2^q$, we ignore calculating $M_{X_1^q}(1, j)$.

The approach is sound that there cannot be any false dismissals. Using this approach, however, we could not guarantee to filter all meaningless forks since it also depends on the online calculated alignment scores.

## 4. REUSING SCORE CALCULATIONS

There might exist duplicated substrings in both $T$ and $P$. This situation is even going further when considering $T$ can reach length of a few gigabytes and $P$ can reach length of several megabytes.

Obviously, we do not want to align a substring of $T$ against a substring of $P$ more than once if they have been aligned before.

The BASIC algorithm in Section 2 represents a distinct substring $X$ starting at positions $t_1, \ldots, t_k$ in $T$ using a suffix path from the root to a leaf node in the suffix trie of $T$. In this way, $X$ is needed to be calculated only once and the alignments $A(t_1 + i, j), \ldots, A(t_k + i, j)$ can share the same alignment score $M_X(i, j)$.

A natural question is whether we could share previous calculated scores on duplicated substrings in $P$ to speed up the alignment process. In this section, we show the techniques of reusing score calculations between forks. We first analyze the relationship between scores of duplicated substrings in $P$. We then show how to identify duplicated substrings in $P$ that can be reused and present an algorithm to reuse score calculations efficiently.

### 4.1 Duplicate Alignment Scores

Consider a matrix $M_X$ shown in Fig. 4. There are two forks starting from $(1, \pi_1)$-entry and $(1, \pi_2)$-entry, respectively. $P_s$ is the common prefix of $P[\pi_1, m]$ and $P[\pi_2, m]$. We mark the entries with common prefix $P_s$ using black color. In the black areas, two alignment scores $M_X(i, \pi_1 + s)$ and $M_X(i, \pi_2 + s)$ are equivalent $(0 \leq s \leq |P_s|)$, since the substring $P[\pi_1, \pi_1 + s]$ equals to $P[\pi_2, \pi_2 + s]$ and they must find the same alignment against $X$.

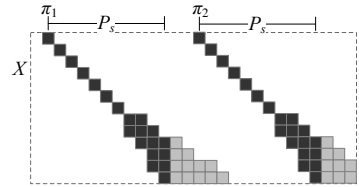

**Figure 4: Entries with a common prefix $P_s$ can share alignment scores. (Black areas represent reusable alignment entries.)**

LEMMA 2. *Given a matrix for a substring $X$. Let $P[\pi_1, \pi_1 + q - 1]$ and $P[\pi_2, \pi_2 + q - 1]$ be two substrings that match $X[1, q]$ exactly. Let $P_s$ be the common prefix of $P[\pi_1, m]$ and $P[\pi_2, m]$. For any two alignment scores $M_X(i, \pi_1 + s)$ and $M_X(i, \pi_2 + s)$, we say $M_X(i, \pi_1 + s)$ equals to $M_X(i, \pi_2 + s)$, if $0 \leq s \leq |P_s|$.*

Fig. 5 shows another case that alignment scores in two forks could be duplicate. Assume the substrings $P[\pi_1, j_1 - 1]$ and $P[\pi_2, j_2 - 1]$ in the figure do not match. Instead, we could find a common substring $P_s$ starting from the two FGOEs. Theorem 5 shows that the two black areas with common substring $P_s$ could share alignment scores if the scores of the two FGOEs are equivalent.

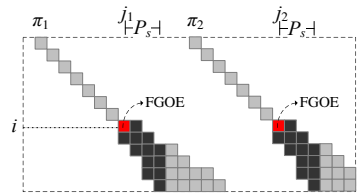

**Figure 5: If two forks have equivalent scores for their FGOEs, their entries with common substring $P_s$ can share alignment scores. (Black areas represent reusable alignment entries.)**

THEOREM 5. *Let $f_1$ and $f_2$ be two forks starting from $(1, \pi_1)$-entry and $(1, \pi_2)$-entry in a matrix $M_X$, respectively. Let $(i_1, j_1)$-entry and $(i_2, j_2)$-entry be the FGOEs of $f_1$ and $f_2$, respectively. If $i_1$ equals to $i_2$, then the alignment score $M_X(i_1, j_1)$ must be equal to the score $M_X(i_2, j_2)$.*

For example, let $X$=GCTACCCCCTTTGGAA, $q$=4, $P[\pi_1, j_1 - 1]$=GCTACACCCTTT and $P[\pi_2, j_2 - 1]$=GCTACCTCCTTT. Since the $q$-prefix $X[1, q]$=GCTA matches $P[\pi_1, \pi_1+q-1]$ and $P[\pi_2, \pi_2+q-1]$, there are two forks starting from $(1, \pi_1)$-entry and $(1, \pi_2)$-entry in a matrix $M_X$, respectively. Although $P[\pi_1, j_1 - 1]$ and $P[\pi_2, j_2 - 1]$ do not match, using Equation 3, their FGOEs have the same score $M_X(12, \pi_1 + 11) = M_X(12, \pi_2 + 11) = 8$.

LEMMA 3. *Suppose there are two fork areas $f_1$ and $f_2$, and their FGOEs are $(i, j_1)$ and $(i, j_2)$ respectively. Let $P_s$ be the common prefix of $P[j_1, m]$ and $P[j_2, m]$, then $\forall 0 \leq s \leq |P_s|$ and $i' \geq i$, $M_X(i', j_1 + s)$ equals to $M_X(i', j_2 + s)$.*

## 4.2 Identify Duplicates in a Query

Reexamine the two cases described in Figs. 4 and 5. The reusable entries belong to two parts: no gap regions and gap regions. If an entry $(i, j)$ belongs to a no gap region, we could use Equation 3 to calculate the score $M_X(i, j)$. If an entry $(i, j)$ belongs to a gap region, however, we have to first calculate the other two auxiliary scores $G_a(i, j)$ and $G_b(i, j)$, and then choose a maximal value among $G_a(i, j)$, $G_b(i, j)$, and the score calculated using Equation 3. It takes more time to calculate a score in a gap region than in a no gap region. Therefore, we focus on reusing scores of entries in gap regions in this section.

Consider a matrix $M_X$ that contains $k$ forks. Let $(i, j_1)$, ..., and $(i, j_k)$ be FGOEs of these forks respectively. When calculating the alignment score of the $(i, j)$-entry in the gap region in a fork with FGOE $(i, j_w)$ ($j_1 \leq j_w \leq j_{k-1}$), we need to record the substring $P[j_w, j]$ so that whenever we meet a duplicate of this substring starting from the next position $j_{w+1}$ in $P$, we could reuse scores of entries in between columns $j_w$ and $j$ to entries in between columns $j_{w+1}$ and $j + j_{w+1} - j_w$. In order to do it, we need to identify duplicates among any two substrings $P[j_u, m]$ and $P[j_v, m]$, where $j_1 \leq j_u, j_v \leq j_k$ and $j_u \neq j_v$.

A straightforward approach is to build up a path for each suffix $P[j_w, m]$. For each node $u$ in the tree, we merge its child nodes if they have the same input edge from $u$. However, this approach is both time and space consuming. We could build up a common prefix tree $T_{P_s}$ in linear time on-the-fly (see Algorithm 2). Instead of process each suffix, we use a set of disjoint substrings $\{P[j_1, j_2 - 1], P[j_2, j_3 - 1], ..., P[j_k, m]\}$ to construct $T_{P_s}$, since each suffix $P[j_w, m]$ can be assembled by concatenating $P[j_w, j_{w+1}-1]$, ..., and $P[j_k, m]$.

The algorithm CONSTRUCTCPTREE first initializes $T_{P_s}$ using a root node $root$ (line 1). It then inserts each substring $S = P[j_i, j_{i+1} - 1]$ into $T_{P_s}$. The algorithm CONSTRUCTCPTREE checks if $root$ of $T_{P_s}$ has an outgoing edge. If there is no outgoing edge from $root$, it directly creates a child node $c$ of $root$ and labels the edge between $root$ and $c$ using $S$. Notice that, $S$ is only the prefix of $P[j_i, m]$. When processing $P[j_{i+1}, j_{i+2} - 1]$, we need to concatenate $S$ to corresponding leaf nodes in $T_{P_s}$. The algorithm sets a link from $root$ to $c$ to mark such a leaf node, which need to be processed for succeeding substring (lines $4 - 5$). Otherwise, the algorithm searches $P[j_i, j_{i+1} - 1]$ in $T_{P_s}$ until it reaches a node $u$ with level $l$ such that the substring $S[1, l]$ is represented by $path(root, u)$ (line 7). If there exists an outgoing edge that can match a substring $S[l + 1, l']$ from $l + 1$ in $S$, then the algorithm splits the edge into two substrings $S[l + 1, l']$ and $S[l' + 1, |S|]$ by inserting a new node $c'$; otherwise it creates a new child node $c$ of $u$ and labels the edge between $u$ and $c$ using $S[l + 1, |S|]$ (lines $8 - 12$). Finally, the algorithm concatenates $S$ to all leaf nodes marked by links (lines $14 - 16$). The algorithm returns the root node of $T_{P_s}$. The time complexity of constructing a common prefix tree for $k$ substrings is $\mathcal{O}(k + \frac{m}{k})$.

**Algorithm 2:** CONSTRUCTCPTREE()

**Input**: A query $P$, a vector of column ids $F_v$;
**Output**: Root of the common prefix tree $T_{P_s}$;

1  Initialize a common prefix tree $T_{P_s}$ using a root node $root$;
2  **foreach** $1 \leq i \leq k - 1$ **do** // process $P[j_w, j_{w+1} - 1]$
3     $S = P[j_w, j_{w+1} - 1]$;
4     **if** *root of $T_{P_s}$ has no outgoing edge* **then**
5        Create a node $c$; $edge(root, c) = S$; $link(root) = c$;
6     $w = link(root)$;
7     Find a deepest node $u$ such that the prefix $S' = S[1, l]$ ($1 \leq l \leq |S|$) can be represented by $path(root, u)$;
8     **if** *there exists a node $v$ such that the prefix of $edge(u, v)$ matches $S[l + 1, l']$ ($l' \leq |S|$)* **then**
9        Split $edge(u, v)$ by inserting a node $c'$;
10       Create a node $c$; $edge(c', c) = S[l' + 1, |S|]$;
11    **else**
12       Create a node $c$; $edge(u, c) = S[l + 1, |S|]$;
13    $link(root) = c$;
14    **while** $w \neq null$ **do**
15       Set a temp link node $v=w$;
16       Create a node $c$; $edge(w, c)=S$; $w = link(v)$; $link(v) = c$;
17    return $root$;

For example, let $P$=CACGTATACG and assume $j_1 = 2, j_2 = 4, j_3 = 6, j_4 = 8$. The constructed procedures for each inserted substring $P[j_i, j_{i+1} - 1]$ ($1 \leq i < 4$) are shown in Fig. 6. Fig. 6(a) shows the tree that only contains a substring $P[j_1, j_2 - 1]$=AC. When inserting the second substring $P[j_2, j_3 - 1]$=GT, the function CONSTRUCTCPTREE creates a node of root and let the edge from root to this created node be GT since GT has no common prefix with substrings in the tree shown in Fig. 6(a). It then processes nodes pointed by links from the root and concatenate GT to the leaf node under AC and modifies links as shown in Fig. 6(b). Fig. 6(c) shows the common prefix tree after inserting $P[j_3, j_4 - 1]$=AT. Since AT and the edge from root to its left child AC has a common prefix A, the function needs to split the edge AC into A and C by inserting a new node. It then concatenates T under the new inserted node and processes all the nodes pointed by links. Fig. 6(d) shows the final common tree for $P[j_1, |P|]$=ACGTATACG, $P[j_2, |P|]$=GTATACG, $P[j_3, |P|]$=ATACG, and $P[j_4, |P|]$=ACG.

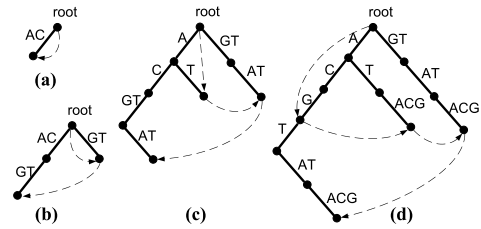

**Figure 6: An example of constructing a common prefix tree.**

Notice that, the online created suffix paths are only related to substrings with the same prefix $X[1, q]$. We could release the online created suffix paths when processing another matrix with different prefix $X'[1, q]$.

### 4.3 A Hybrid Algorithm for Efficiently Reusing Score Calculations

In order to reuse the alignment scores of entries in one fork, instead of calculating the matrix row by row, we do it in hybrid. That is, we calculate scores of entries in no gap regions horizontally to identify FGOEs with the same row, we then calculate scores of

**Algorithm 3:** HYBRID - reusing score calculations

**Input**: A substring $X$, a query pattern $P$, alignments $A$;
**Output**: The matrix $M_X$;
1 Find positions $pos_{set} = \{t_1, \ldots, t_k\}$ of all the occurrences of $X[1, q]$ in $P$;
2 Initialize a matrix $M_X$ using a hash table;
   // process entries in no gap regions
3 $F_{set} = calMatrixByRow(X, P, q, pos_{set}, M_X)$;
4 **while** $F_{set} \neq \emptyset$ **do** // process entries in gap regions
5     Pop all FGOEs with the same row id from $F_{set}$ and push the correponding column ids into a vector $F_v$;
6     $calMatrixByColumn(X, P, M_X, F_v)$;
7 return $M_X$;

**Function** $calMatrixByRow$

**Input**: A substring $X$, a query $P$, an integer $q$, starting positions $pos_{set} = \{t_1, \ldots, t_k\}$ of occurrences of $X[1, q]$ in $P$, a matrix $M_X$;
**Output**: A queue of FGOEs $F_{set}$;
1 Initialize a queue $F_{set}$ to store FGOEs of forks in $M_X$;
2 **foreach** $1 \leq i \leq L_{max}$ **do** // length filtering
3    **if** $pos_{set} == \emptyset$ **then**
4      break;
5    **foreach** position $t$ in $pos_{set}$ **do**
6      **if** $i \leq q$ **then**
7        $M_X(i, t+i-1) = s_a \times i$;
8      **else**
9        $M_X(i, t+i-1) = M(i-1, t+i-2) + \delta(X[i], P[t+i-1])$;
10      **if** $M_X(i, t+i-1) > |s_g + s_s|$ **then**
11        $F_{set}$.push_back$(i, t+i-1)$;
12        $pos_{set} = pos_{set} - \{t\}$;
13      **if** $M_X(i, t+i-1)$ *does not satisfy score filtering* **then**
14        $pos_{set} = pos_{set} - \{t\}$;
15 return $F_{set}$;

---

entries in gap regions vertically to make copy operations between forks efficiently. Algorithm 3 is an overview of the HYBRID algorithm. The HYBRID algorithm first locates positions of all matching grams of $X[1, q]$ in $P$ using the inverted lists of $q$-grams for $P$ (line 1). It then initializes a hash table to store entries in a matrix $M_X$ for the substring $X$ and the query $P$.

In order to calculate all FGOEs in $M_X$, it invokes a function $calMatrixByRow$ to calculate scores of entries row by row in no gap regions until all FGOEs have been found. The FGOEs are pushed into a queue $F_{set}$ (line 3). Based on the calculated FGOEs, the HYBRID algorithm processes entries in each gap region represented by an FGOE column by column. According to Lemma 3, only when the FGOEs have the same row ids, their gap regions could reuse alignment scores. The algorithm HYBRID, therefore, popps all FGOEs with the same row id from $F_{set}$ and pushes them into a vector of FGOEs $F_v$. It then invokes a function $calMatrixByColumn$ to reuse scores of entries in gap regions until all FGOEs in $F_{set}$ have been processed (lines 4 – 6). It finally returns the matrix $M_X$ represented by a hash table (line 7).

**Horizontal calculations in no gap regions**. We use the function $calMatrixByRow$ to show details of calculations in no gap regions. It initializes a queue of FGOEs $F_{set}$. According to length filtering (see Theorem 1), it only processes rows less than or equal to $L_{max}$. For each row id $i$ ($1 \leq i \leq q$), it assigns $s_a \times i$ to $M_X(i, t+i-1)$, where $t \in pos_{set}$ is the starting position of an occurrence of $X[1, q]$ in $P$ (lines 6 – 7). These scores could be assigned without any calculation according to the $q$-prefix filtering in Section 3.1.3. For each row id $i(> q)$ in no gap regions, it calculate scores using Equation 3 (line 9). Notice that, the function $calMatrixByRow$ only needs to calculate scores of entries $(i, t+i-1)$ ($t_1 \leq t \leq t_k$) since only these entries might be meaningful. If a score $M_X(i, t+i-1)$ is greater than $|s_g + s_s|$, it means the corresponding $(i, t+i-1)$-entry is an FGOE (see Section 3.1.3). The function pushes the entry $(i, t+i-1)$ into the queue $F_{set}$ and removes the starting position $t$ from $pos_{set}$ (lines 10 – 12). If the score $M_X(i, t+i-1)$ does not satisfy score filtering (see Theorem 2), the function removes $t$ from $pos_{set}$ (lines 13 – 14). We repeat the above process until $pos_{set}$ becomes empty.

**Vertical calculations in gap regions**. In order to reuse scores of duplicates in $P$ efficiently, we identify duplicates and reuse alignment scores among gap regions. As the Algorithm HYBRID shows, for each iteration, the function $calMatrixByColumn$ processes a gap region starting from each position in $F_v$. It first constructs a common prefix tree $T_{P_s}$ to identify duplicate substrings using $P$ and $F_v$ (line 1). For each substring $P[j_w, j_{w+1} - 1]$, the function does not calculate the score of an entry $(i, j)$ unless it could not find a duplicate in the common prefix tree $T_{P_s}$ (lines 2 – 21).

Before calculating score $M_X(i, j)$ in a gap region, the function $calMatrixByColumn$ checks if a prefix of $P[j_w, j_{w+1} - 1]$ is a duplicate of another substring starting at a previous position $j_h$ ($j_1 \leq j_h < j_w$) in $F_v$. It finds a path $path(r, z)$ to represent $P[j_w, j_{w+1}-1]$ (line 3) and then evaluates each edge in $path(r, z)$ (line 5). While an edge $edge(u, v)$ has been processed, the function could reuse entries associated with this edge, i.e. it copies each score $M_X(i, v.column + d)$ to $M_X(i, j_w + d)$ (lines 7 – 9). The function $calMatrixByColumn$ repeats the above iteration until it meets an edge that has not been processed. For each remaining edge in $path(r, z)$, it calculates a range of row ids $[s, e]$ such that for each row id $i \in [s, e]$ the $(i, j_w + d)$-entry belongs to the current processing gap region. It calculates the alignment scores for each $(i, j_w + d)$-entry and records this range $[s, e]$ so that other gap regions could reuse scores of entries in this column (lines 14 – 18). Notice that, when calculating $M_X(i, j_w + d)$ in a gap region, the function takes the advantage of our vertical calculation. It only needs one byte to store the auxiliary score $G_a(i-1, j_w + d)$ and a vector to store the auxiliary score $G_b$ in the $(j_w+d-1)$-th column. The size of the vector equals to $e - s + 1$. These auxiliary scores could be released after calculating $M_X(i, j_w + d)$. The calculation stops when all the entries in the $(j_w + d)$-th column of the current gap region are meaningless. It releases $T_{P_s}$ after all positions in $F_v$ have been processed since $T_{P_s}$ is only used locally (line 22).

**Reuse substrings in text $T$ with the same prefix $X[1, q]$**. In the BASIC algorithm, the suffix trie $T$ of $T$ provides an advantage of avoiding aligning substrings of $T$ that are identical. When moving from a node in $T$ to its child node, a row is added to the calculation matrix and when a node is going up to its father node, the last row is deleted. Hence, the common prefixes in $T$ are only aligned once.

In order to combine the above advantage in our hybrid calculations, we process paths in $T$ which have the common prefix $X[1, q]$. For each such path, we identify the forks whose FGOEs no longer exist as we are going up along the suffix trie and recalculate the FGOEs for them using the function $calMatrixByRow$. Then, we vertically calculate these newly updated gap regions using the function $calMatrixByColumn$. We repeat the above process until all paths with the same $X[1, q]$ prefix have been processed.

```
Function calMatrixByColumn
    Input: A substring X, a query P, a matrix M_X, a vector of column
           ids F_v;
 1  r = u = CONSTRUCTCPTREE(P, F_v);
 2  foreach 1 ≤ w ≤ k − 1 do // process P[j_w, j_{w+1} − 1]
 3      Find a node z such that P[j_w, j_{w+1} − 1] can be represented by
          path(r, z);
 4      Let v be the child node of u in path(r, z);
        // Reuse entries in gap regions
 5      while v.column > 0 do // the substring
          represented by edge(u, v) has been processed
 6          d = 0;
 7          while d < edge(u, v).length() do
              // calculate the (j_w + d)-th column
 8              for i ∈ v.range[j_w + d] do
 9                  M_X(i, j_w + d) = M_X(i, v.column + d);
10              d + +;
11          u = v; v = the child node of u in path(r, z);
12      repeat // Calculate entries in gap regions
13          d = 0;
14          while d < edge(u, v).length() do
15              Calculate a range of row ids [s, e] such that ∀i ∈ [s, e],
                  the (i, j_w + d)-entry belongs to the current gap region;
16              for i ∈ [s, e] do
17                  Calculate M_X(i, j_w + d);
18              v.range[j_w + d] = [s, e];
19          v.column = j_w;
20          u = v; v = the child node of u in path(r, z);
21      until all the scores in the (j_w + d)-th column of the current gap
          region are meaningless;
22  Release the common prefix tree T_{P_s} rooted at r;
```

## 5. SIMULATING SEARCHES USING COMPRESSED SUFFIX ARRAY

As discussed in Sections 3 and 4, our technique requires the following three kinds of searches in the suffix trie $\mathcal{T}$ of a text $T$:

*(1) Given a q-length substring $S^q$ in the query $P$, check if $S^q$ appears in $T$ exactly.* In the scenario of this paper, for each suffix path in $\mathcal{T}$, the alignment scores of the $(i + 1)$-th row in the matrix $M$ depend on the scores of the $i$-th row in $M$. Therefore, based on the $SA$ range of $X$, we hope to process the string $X' = Xc$ by iteratively appending one character $c$ behind $X$. We use the technique reported in [8] to construct a compressed suffix array for the reversal of $T$ (denoted $T^{-1}$). In $T^{-1}$, we do not change the position of each character in $T$ and let position of $ be 0.

Given a $q$-length substring $S^q$, we search $(S^q)^{-1}$ using the backward search algorithm [6] on the compressed suffix array and get an $SA$ range from $SA[i]$ to $SA[j]$ for $(S^q)^{-1}$. $S^q$ does not appear in $T$ only when $i < j$. This search operation could be done in $\mathcal{O}(q)$ steps and each step costs constant time.

*(2) Given a substring $X_s = X[1, i]$, find its starting positions of all occurrences in $T$.* We search $X_s^{-1}$ and find the $SA$ range from $SA[x]$ to $SA[y]$ for $X_s^{-1}$. The position of each appearance of $X_s$ in $T$ is $SA[h] − |X_s| + 1$, where $x \leq h \leq y$. For example, $T$ = GCTAGC$, and $T^{-1}$ = C⁶G⁵A⁴T³C²G¹$⁰, where integers represent positions of characters in $T$. Let $|X_s|$ be GC, we search its reversed string CG and get the $SA$ range $[2, 3]$. The suffix array $SA[0, 6] = \{0, 4, 2, 6, 1, 5, 3\}$, so the positions of the query GC are $SA[2] − |X_s| + 1 = 1$ and $SA[3] − |X_s| + 1 = 5$.

Notice that, in the matrix $M_X$, we always process $X[1, i]$ after $X[1, i − 1]$. Since $X[1, i]$ equates to appending $X[i]$ behind $X[1, i − 1]$, we could find the appearances of $X[1, i]$ in $T$ in $\mathcal{O}(1)$ time based on the $SA$ range of $X[1, i − 1]$ using the backward search algorithm [6].

*(3) Given a q-prefix $X^q$, traverse the suffix trie and get suffix paths whose represented substrings have the same prefix as $X^q$.* Let $u$ be a node in the conceptual suffix trie and $X^q$ be the represented substring of the path from the root to $u$.

Since we have found the $SA$ range of $(X^q)^{-1}$ using the compressed suffix array of $T^{-1}$, we can check the existence of edge with label $c$ from $u$ by computing the $SA$ range for $c(X^q)^{-1}$. We enumerate the corresponding substring if the edge $c$ does exist and repeat the same procedure to traverse the subtree rooted at $u$.

## 6. ANALYSIS OF NUMBER OF CALCULATED ENTRIES

We consider the general scoring scheme $\langle s_a, s_b, s_g, s_s \rangle$. As we have analyzed in Section 3.1, the larger the $\frac{|s_b|}{|s_a|}$, $\frac{|s_g|}{|s_a|}$, and $\frac{|s_s|}{|s_a|}$ are, the better performance of local filtering techniques could be. In order to understand the behaviors of ALAE deeply, we analyze the number of calculated entries.

We consider the general scoring scheme for any random substring $X^d$ with $d$ characters ($d \geq 1$) in the text $T$. According to score filtering, we are interested in each alignment substring $P'$ of the query $P$ such that the alignment scores between $P'$ and $T$ are positive. For a simple case that an alignment cannot insert a space or a gap, let $P'$ contain $d$ characters. We define $f(d)$ to be the number of length-$(d)$ substring $P'$ such that $score(X^d, P') > 0$.

LEMMA 4. *When there is no gap between $X^d$ and $P'$, we have $f(d) \leq k_1(k_2)^d$, where $k_1 = (1 − \frac{1}{s})^q (\frac{\sigma-1}{\sigma-2}) \frac{s}{\sqrt{2\pi(s-1)}}$, $k_2 = s \sqrt[s]{\frac{\sigma-1}{(s-1)^{s-1}}}$, and $s = 1 + \frac{|s_b|}{|s_a|}$.*

PROOF. Since $s_a > 0$ and $s_b < 0$, when $score(X^d, P') > 0$, the largest number of mismatches is $\lfloor d/s \rfloor$. According to prefix filtering in Theorem 3, mismatches could not appear in either $X[1, q]$ or $P'[1, q]$.

Therefore, $f(d) \leq \sum_{i=0}^{\lfloor \frac{d}{s} \rfloor} (\sigma − 1)^i \binom{d-q}{i}$. Since $\binom{d}{i} = \frac{d}{d-i} \binom{d-1}{i}$, we conclude $\binom{d-q}{i} = \frac{(d-i)(d-1-i)...(d-(q-1)-i)}{d(d-1)...(d-(q-1))} \binom{d}{i} \leq (1 − \frac{i}{d})^q \binom{d}{i}$, then $f(d) \leq \sum_{i=0}^{\lfloor \frac{d}{s} \rfloor} (\sigma-1)^i (1-\frac{i}{d})^q \binom{d}{i} \leq (\frac{\sigma-1}{\sigma-2})(\sigma-1)^{\frac{d}{s}}(1-\frac{1}{s})^q \binom{d}{\lfloor \frac{s}{d} \rfloor}$.

As we know, $\binom{d}{i} = \frac{d!}{(d-i)!i!}$. Using the Stirling's approximation, $d! = \sqrt{2\pi d}(\frac{d}{e})^d e^{\lambda_d}$, where $\frac{1}{12d+1} < \lambda_d < \frac{1}{12d}$. Therefore,

$$\frac{d!}{(d-i)!i!} = \frac{\sqrt{2\pi d}(\frac{d}{e})^d e^{\lambda_d}}{\sqrt{2\pi(d-i)}(\frac{d-i}{e})^{d-i}e^{\lambda_{d-i}}\sqrt{2\pi d}(\frac{i}{e})^i e^{\lambda_i}}$$

$$= \frac{d^{d+\frac{1}{2}}}{\sqrt{2\pi}(d-i)^{d-i+\frac{1}{2}}(i)^{i+\frac{1}{2}}} e^{\lambda_d - \lambda_{d-i} - \lambda_i}.$$

Obviously, $\lambda_d − \lambda_{d-i} − \lambda_i \leq 0$, then $e^{\lambda_d - \lambda_{d-i} - \lambda_i} \leq 1$. Thus,

$$\binom{d}{i} \leq \frac{d^{d+\frac{1}{2}}}{\sqrt{2\pi}(d-i)^{d-i+\frac{1}{2}}(i)^{i+\frac{1}{2}}} = \frac{1}{\sqrt{2\pi i}} (\frac{d}{d-i})^{d+\frac{1}{2}} (\frac{d-i}{i})^i.$$

So, $\binom{d}{\lfloor \frac{d}{s} \rfloor} \leq \frac{\sqrt{s}}{\sqrt{2\pi d}}(\frac{d}{d-\frac{d}{s}})^{d+\frac{1}{2}}(\frac{d-\frac{d}{s}}{\frac{d}{s}})^{\frac{d}{s}} = \frac{s}{\sqrt{2\pi(s-1)d}}(\frac{s}{(s-1)^{\frac{s-1}{s}}})^d$
$\leq \frac{s}{\sqrt{2\pi(s-1)}}(\frac{s}{(s-1)^{\frac{s-1}{s}}})^d$. Therefore,
$f(d) \leq (1-\frac{1}{s})^q (\frac{\sigma-1}{\sigma-2}) \frac{s}{\sqrt{2\pi(s-1)}} \left(s \sqrt[s]{\frac{\sigma-1}{(s-1)^{s-1}}}\right)^d = k_1(k_2)^d$. □

Based on the analysis in [8], the expected total number of calculated entries of ALAE is

$$(\frac{k_1}{k_2 - 1} + \frac{k_1 \sigma^2}{\sigma - k_2}) mn^{\log_\sigma k_2}. \tag{4}$$

BLAST specifies scoring parameters at http://blast.ncbi.nlm.nih.gov/Blast.cgi, where $(s_a, s_b) \in \{(1, −2), (1, −3), (1, −4), (2, −3),$

$(4, -5), (1, -1)\}$. For most of the parameters, $\frac{|s_g|}{|s_a|} \in \{1, 2, 3, 5\}$ and $\frac{|s_s|}{|s_a|} \in \{1, 2\}$. According to Equation 4, the upper bound of the number of calculated entries for DNA sequences can vary from $4.50mn^{0.520}$ to $9.05mn^{0.896}$, and for protein sequences can vary from $8.28mn^{0.364}$ to $7.49mn^{0.723}$. Notice that both BLAST and BWT-SW use $\langle 1, -3, -5, -2 \rangle$ as the default scoring scheme when aligning DNA sequences. Using this scoring scheme, the number of calculated entries using BWT-SW is upper bounded by $69mn^{0.628}$ (see [8]), whereas using ALAE the number is upper bounded by $4.47mn^{0.6038}$.

## 7. EXPERIMENTS

In this section, we report our experimental results of ALAE.
**Data sets:** We used the following commonly used real data sets, including two DNA data sets and one protein data set. The alphabet size of DNA sequences and protein sequences are 4 and 20, respectively.

*Human genome data set.* The human reference sequence (GRCh 37) was assembled from a collection of DNA sequences.[2] It consists of 24 chromosomes ranging in length from 48 million to 249 million. We used subsequences with different lengths of GRCh37 as texts to be aligned with, and the lengths varied from 50 million to 1 billion.

*Mouse genome data set.* The mouse genome (MGSCv37 chr1) was extracted from house mouse that contains 198 million characters.[3] Since aligning mouse genomes against human genomes is widely used to do homology search in practice [7, 12], we used MGSCv37 chr1 to generate queries against human genomes. We randomly chose 100 starting positions in the first 180 million characters and picked a fixed length substring from each randomly located starting position to generate a query workload that contains 100 query sequences with the same query length. We varied the query length from 1 thousand to 1 million. We used these query workloads to test performance of ALAE.

*Protein data set.* We used the comprehensive and non-redundant database UniParc[4] that contains most of the publicly available protein sequences in the world. We varied the lengths of texts (protein sequences) from 10 million to 50 million. We randomly chose sequences from UniParc as queries, ranging in length from 200 to 100, 000.

**Threshold $H$ and $E$-value:** In our experiments, instead of setting a threshold value $H$ explicitly, we used an Expectation value (a.k.a. $E$-value) that is widely adopted by the biological community. The $E$-value is a parameter that describes the number of alignments one can "expect" to see by chance when searching a database of a particular size. The following equation relates $E$-value and alignment score $S$: $E = Kmne^{-\lambda S}$, where $K$ and $\lambda$ are scaling constants computed by BLAST [1]. The corresponding threshold $H$ for ALAE can be computed as follows [11]: $H = \lceil \frac{ln(Kmn) - ln(E)}{\lambda} \rceil$. We varied the $E$-value from $10^{-15}$ to 10. Both BLAST and BWT-SW set $E = 10$ as the default parameter.

**Scoring scheme:** In our experiments, we used the same scoring parameters as BLAST (see Section 6) to evaluate the performance of ALAE. Both BLAST and BWT-SW adopt $\langle 1, -3, -5, -2 \rangle$ as the default scoring scheme. Notice that BWT-SW requires that $|s_b| \geq 3|s_a|$, which highly limits its usability.

[2] http://hgdownload.cse.ucsc.edu/goldenPath/hg18/chromosomes/
[3] http://hgdownload.cse.ucsc.edu/goldenPath/mm9/chromosomes/
[4] ftp://ftp.uniprot.org/pub/databases/uniprot/current_release/uniparc/uniparc_active.fasta.gz

**Table 2: Comparison of alignment time and number of alignment results when varying lengths of queries ($n = 1$ billion).**

| Approaches | Alignment time (Sec.) and number of alignment results $\mathcal{C}$ | | | | | | | | | |
|---|---|---|---|---|---|---|---|---|---|---|
| | $m$=1K | | $m$=10K | | $m$=100K | | $m$=1M | | $m$=10M | |
| | Time | $\mathcal{C}$ | Time | $\mathcal{C}$ | Time | $\mathcal{C}$ | Time | $\mathcal{C}$ | Time | $\mathcal{C}$ |
| ALAE | 0.006 | 994 | 0.080 | 7790 | 1.484 | 34911 | 19.269 | 150390 | 393.001 | 586521 |
| BLAST | 0.033 | 744 | 0.312 | 5928 | 3.074 | 24154 | 31.459 | 99916 | 330.330 | 395652 |
| BWT-SW | 2.144 | 994 | 21.756 | 7790 | 177.048 | 34911 | 1451.448 | 150390 | - | - |

**Table 3: Comparison of alignment time and number of alignment results when varying lengths of texts ($m = 1$ million).**

| Approaches | Alignment time (Sec.) and number of alignment results $\mathcal{C}$ | | | | | | | | | |
|---|---|---|---|---|---|---|---|---|---|---|
| | $n$=50M | | $n$=100M | | $n$=200M | | $n$=500M | | $n$=1G | |
| | Time | $\mathcal{C}$ | Time | $\mathcal{C}$ | Time | $\mathcal{C}$ | Time | $\mathcal{C}$ | Time | $\mathcal{C}$ |
| ALAE | 5.272 | 21172 | 6.288 | 82702 | 6.537 | 100562 | 12.702 | 114691 | 19.269 | 150390 |
| BLAST | 18.489 | 13290 | 21.362 | 56377 | 22.160 | 83836 | 28.669 | 90864 | 31.459 | 99916 |
| BWT-SW | 84.827 | 21172 | 147.969 | 82702 | 235.236 | 100562 | 617.452 | 114691 | 1451.448 | 150390 |

All the algorithms were implemented using GNU C++. The experiments were run on a PC with an Intel 2.93GHz Quad Core CPU i7 and 8GB memory with a 500GB disk, running a Ubuntu (Linux) operating system.

### 7.1 Alignment Time and Number of Results

We compared ALAE with three state-of-art algorithms: Smith-Waterman algorithm, BLAST, and BWT-SW[5] using the default settings of both BLAST and BWT-SW (i.e. $\langle 1, -3, -5, -2 \rangle$ and $E = 10$). We would not include the Smith-Waterman algorithm into our following discussions because this algorithm is too slow to be considered. Our experiments show that the Smith-Waterman algorithm took 7.7 hours to align a query with 10 thousand characters against a text with 50 million characters. However, ALAE only took 25 ms. For the same reason, we would not report the query performance for the BASIC algorithm since it has higher time complexity than the Smith-Waterman algorithm. We conducted experiments to show the average time required by ALAE, BLAST, and BWT-SW under different circumstances.

Table 2 shows the average alignment time and the number of alignment results when varying the lengths of queries from 1 thousand to 10 million. We used a 1 billion human genome sequence as the text. ALAE shows a great advantage over BWT-SW with all queries and can find the results as BWT-SW does. Notice that our experiments show that BWT-SW could not align a query with more than 1 million characters against the text due to insufficient memory. ALAE outperformed BLAST when the query lengths were less than 10 million. However, when the query length was extremely long, such as 10 million, the alignment time was not as fast as BLAST. It is worth mentioning that ALAE found more results than BLAST did.

Table 3 shows the average alignment time and number of alignment results when varying the length of a text from 50 million to 1 billion. We used the query workload, in which each query had 1 million characters. Table 3 shows ALAE outperforms both BWT-SW and BLAST for different text lengths when $m = 1$ million.

We also conducted experiments on protein sequences. The results for the protein data sets are similar to those presented here. For space reason, we omit these results in this paper.

### 7.2 Filtering Ratio and Reusing Ratio

In this section, we show the effectiveness of our proposed filtering and entry reusing techniques. We use *filtering ratio* to evaluate our filtering techniques compared with BWT-SW:

$$\text{Filtering ratio} = \frac{\text{\# of filtered entries}}{\text{\# of calculated entries using BWT-SW}} \times 100\% \quad (5)$$

[5] Available at http://i.cs.hku.hk/ ckwong3/bwtsw

**Table 4: Number of calculated entries and their computation costs** ($n = 1$ **billion, score =** $\langle 1, -3, -5, -2 \rangle$**).**

| Approaches | $m = 10,000$ | | $m = 100,000$ | | $m = 1,000,000$ | |
|---|---|---|---|---|---|---|
| | # of calculated entries × cost | Computation cost | # of calculated entries × cost | Computation cost | # of calculated entries × cost | Computation cost |
| ALAE | 31,865 × 1<br>103,403 × 2<br>329,124 × 3 | 1,226,043 | 318,640 × 1<br>2,139,094 × 2<br>7,009,018 × 3 | 25,623,882 | 3,266,537 × 1<br>25,890,567 × 2<br>88,159,153 × 3 | 319,525,130 |
| BWT-SW | 1,245,288 × 3 | 3,735,864 | 21,827,128 × 3 | 65,481,384 | 271,024,617 × 3 | 813,073,851 |

Filtered entries are the entries which are calculated using BWT-SW but are considered meaningless using ALAE. A higher filtering ratio means our filtering techniques are more effective.

We use *reusing ratio* to evaluate our entry reusing technique:

$$\text{Reusing ratio} = \frac{\text{\# of reused entries}}{\text{\# of accessed entries}} \times 100\% \quad (6)$$

Reused entries are the ones whose scores can be simply copied from previous calculated entries using ALAE. Accessed entries consist of both reused entries and calculated entries. The higher the reusing ratio is, the more effective our reusing technique is.

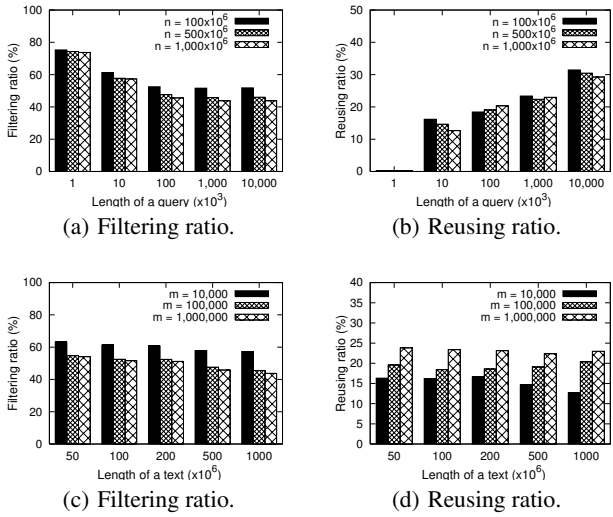

**Figure 7: Filtering and reusing ratios using** $\langle 1, -3, -5, -2 \rangle$**.**

We conducted experiments with different lengths of queries and texts. Figs. 7 shows the filtering ratios and reusing ratios using ALAE, when $E = 10$ and the scoring scheme is $\langle 1, -3, -5, -2 \rangle$.

Fig. 7(a) shows that a query workload with shorter queries maintains a higher filtering ratio. For a fixed text with 100 million characters, the filtering ratio decreased from 75.3% to 51.8% when the query length increased from 1 thousand to 10 million. The reason is that when the query length is short, calculated entries are mainly belonging to no gap regions. In a no gap region, ALAE could filter more meaningless entries compared with BWT-SW.

Fig. 7(b) shows when the query length increases from 10 thousand to 10 million, the reusing ratio increases from 16.2% to 31.5%. This is because longer queries contain more repetitions and more entries can be reused during the alignment process. When the query length is 1 thousand, the reusing ratio is very low since it is hard to find duplicates among forks.

Figs. 7(c) and 7(d) show both the filtering ratio and reusing ratio keep stable when changing the length of a text for a fixed query workload. The reason is that ALAE only considers a substring with small length $L_{max}$ to do alignment against the query.

ALAE not only reduces the number of entries that need to be calculated by BWT-SW, but also optimizes the cost for computing scores. Table 4 shows how the number of calculated entries translates into the computation cost. Using ALAE, calculated entries could belong to a no gap region or a gap region in each matrix. In a no gap region, ALAE uses the simplified recurrence function (see Equation 3) to calculate score for each entry, whereas BWT-SW has to consider the auxiliary scores in both $G_a$ and $G_b$, which requires extra computation costs. Similarly, the entries in the boundaries of the fork areas using ALAE only rely on their two adjacent entries instead of three adjacent entries using BWT-SW.

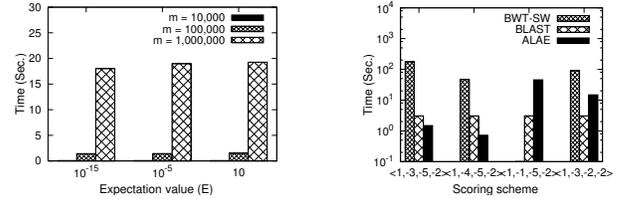

**Figure 8: Varying $E$-values under $\langle 1, -3, -5, -2 \rangle$.**   **Figure 9: Varying scoring schemes ($E=10$, $m=100,000$).**

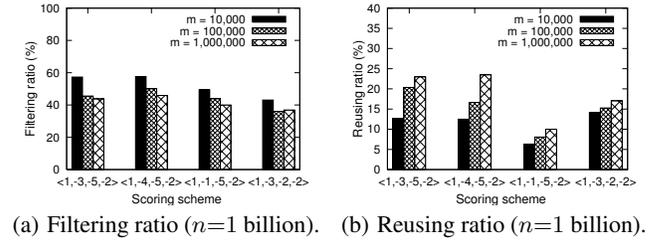

(a) Filtering ratio ($n$=1 billion).   (b) Reusing ratio ($n$=1 billion).

**Figure 10: Filtering and reusing ratios using different scores.**

### 7.3 Effect of $E$-Values

In this section, we examine how ALAE would be affected by $E$-values. Fig. 8 shows the alignment time of ALAE when varying $E$ from $10^{-15}$ to 10 using three different query workloads. We can see that ALAE is not very sensitive to $E$-values. For the query workload that contains queries with 10,000 characters, the alignment time is 72ms when $E$ is $10^{-15}$, 72.9ms when $E$ is $10^{-5}$, and 79.9ms when $E$ is 10. The time is too small to be seen in this figure. For any given query workload, ALAE shows small time rises when we increase $E$. The reason of these time rises is that a large $H$ value (i.e. small $E$ value) terminates calculations earlier than a small $H$ value. Notice that such rises are very small since score filtering only shares a small impact on accelerating alignment time.

### 7.4 Effect of Scoring Schemes

In order to test the effect of scoring schemes, we chose four scoring schemes in BLAST by varying values $s_a$, $s_b$, $s_g$, and $s_s$.

Fig. 9 shows the four representative scoring schemes that cover large and small values of $q$ and $|s_g| + |s_s|$. Both ALAE and BWT-SW are sensitive to scoring schemes, whereas BLAST is barely affected by the change of scoring schemes, because BLAST adopts a different heuristic approach to find results.

ALAE runs much faster than BWT-SW for all scoring schemes. Notice that we do not include the result of BWT-SW for $\langle 1, -1, -5, -2 \rangle$ since BWT-SW requires that $|s_b| \geq 3|s_a|$. ALAE shows good performance when the scoring scheme is $\langle 1, -3, -5, -2 \rangle$ or $\langle 1, -4, -5, -2 \rangle$. ALAE is 119 times faster than BWT-SW using $\langle 1, -3, -5, -2 \rangle$ and 65 times faster using $\langle 1, -4, -5, -2 \rangle$. This is because a larger $\frac{|s_s|}{|s_a|}$ makes $L_{max}$ much tighter, a smaller $s_a$ makes calculation terminating earlier, and a larger $q$ value (see Equation 2) makes prefix filtering more effective.

Table 5: Number of entries using ALAE.

| Scoring schemes | # of reused entries | # of accessed entries | # of calculated entries |
|---|---|---|---|
| $\langle 1, -1, -5, -2 \rangle$ | 30,652,400 | 380,960,680 | 350,308,280 |
| $\langle 1, -3, -2, -2 \rangle$ | 19,047,958 | 124,804,117 | 105,756,159 |

Fig. 9 shows that ALAE is slower than BLAST when the scoring scheme is $\langle 1, -1, -5, -2 \rangle$. We use Fig. 10 to explain the reason. The small $s_b$ value makes gap regions expanded, which results in large number of calculated entries. Table 5 shows that the number of calculated entries using $\langle 1, -1, -5, -2 \rangle$ is 37 times of the number using $\langle 1, -3, -5, -2 \rangle$. We can see that the reusing ratio is much lower than other scoring schemes. This result is consistent with the analysis in Section 6, where $\langle 1, -1, -5, -2 \rangle$ corresponds with the worst case where the number of calculated entries is upper bounded by $9.05mn^{0.896}$. For $\langle 1, -3, -2, -2 \rangle$, the smaller $|s_g| + |s_s|$ value makes the no gap regions smaller, which weakens the effect of filtering techniques (see Fig. 10(a)).

## 7.5 Index Size

We have evaluated the space efficiency of ALAE for both DNA sequences and protein sequences. We varied the length of texts and collected their index sizes. We used the following scoring schemes: $\langle 1, -3, -5, -2 \rangle$ for DNA sequences, and $\langle 1, -3, -11, -1 \rangle$ for protein sequences. Fig. 11(a) shows the index sizes of ALAE for DNA sequences. The alphabet size of the DNA sequences is 4, thus every character in BWT sequence can be stored using 2 bits. As we can see in Fig. 11(a), the indexes for storing dominate relationships of DNA sequences are mostly too small to be seen.

Figs. 11(b) shows the index sizes of ALAE for protein sequences. For relatively small texts, the index for storing dominate relationships is large compared with BWT index. However, as the size of a text grows, the size of the dominate index decreases quickly. The dominate index size is 98.09MB for a text with 10 million characters and decreases to 8.83MB when the length of the text increases to 20 million.

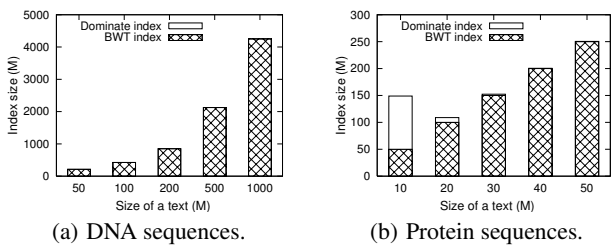

(a) DNA sequences.  (b) Protein sequences.

Figure 11: Index size.

## 8. CONCLUSION AND FUTURE WORK

We have developed a novel approach ALAE to accelerate dynamic programming for finding all local alignments. We gave a full analysis of the dynamic programming approach, and presented a series of filtering techniques to prune meaningless entries and an algorithm to reuse duplicate calculations. Our extensive experiments on real biosequences showed the high efficiency of our techniques. ALAE improves the time efficiency of the state-of-the-art exact BWT-SW approach significantly and accelerates BLAST for most of the scoring schemes. As parts of future work, we will investigate techniques to further improve the performance of ALAE for all scoring schemes and exploit algorithms using external memory.

## 9. ACKNOWLEDGMENTS

The work is partially supported by the National Natural Science Foundation of China (Nos. 60973018, 60973020), the Joint Research Fund for Overseas Natural Science of China (No. 61129002), the National Basic Research Program of China (973 Program) (No. 2012CB316201), the National Natural Science of China Key Program (No. 60933001), the National Natural Science Foundation for Distinguished Young Scholars (No. 61025007), the Doctoral Fund of Ministry of Education of China (No. 20110042110028), and the Fundamental Research Funds for the Central Universities (No. N110804002).